\documentclass[nonacm,sigconf]{acmart}
\setcitestyle{square}
\usepackage{verbatim}
\usepackage{caption}
\usepackage{graphicx}
\usepackage{float} 
\usepackage{subfigure}
\usepackage{subcaption}

\AtBeginDocument{%
  }

\setcopyright{acmlicensed}
\copyrightyear{2018}
\acmYear{2018}
\acmDOI{XXXXXXX.XXXXXXX}

\acmConference[Conference acronym 'XX]{Make sure to enter the correct
  conference title from your rights confirmation emai}{June 03--05,
  2018}{Woodstock, NY}
\acmISBN{978-1-4503-XXXX-X/18/06}

\setcopyright{none}
\renewcommand\footnotetextcopyrightpermission[1]{} 
\begin{document}

\title{Feeling the Grass Grow: Making Midair Haptic Parameters Visible, Touchable and Controllable}


\author{Mingxin Zhang}
 \affiliation{
   \institution{The University of Tokyo}
   \city{Chiba}
   \country{Japan}
 }
\email{m.zhang@hapis.k.u-tokyo.ac.jp}

\author{Qirong Zhu}
 \affiliation{
   \institution{The University of Tokyo}
   \city{Chiba}
   \country{Japan}
 }
\email{e.zhu@hapis.k.u-tokyo.ac.jp}

\author{Yasutoshi Makino}
 \affiliation{
   \institution{The University of Tokyo}
   \city{Chiba}
   \country{Japan}
 }
\email{yasutoshi_makino@k.u-tokyo.ac.jp}

\author{Hiroyuki Shinoda}
 \affiliation{
   \institution{The University of Tokyo}
   \city{Chiba}
   \country{Japan}
 }
\email{hiroyuki_shinoda@k.u-tokyo.ac.jp}
 
\renewcommand{\shortauthors}{Mingxin Zhang, Qirong Zhu, Yasutoshi Makino, and Hiroyuki Shinoda.}


\renewcommand{\abstractname}{\textsc{ABSTRACT}}
\begin{abstract}
In this paper, we present an ultrasound mid-air haptic interaction system that integrates a designed visualization of haptic parameters while maintaining ease of control. The design of corresponding haptic parameters for real-world tactile textures is a complex task. Furthermore, users often face difficulties in simultaneously controlling multi-dimensional haptic parameters to achieve the desired vibration feedback. To address these challenges, the SLS optimization method facilitates user control of these multi-dimensional parameters through a simple one-dimensional slider. Concurrently, our system employs the "Growing Grass" metaphor to visualize haptic parameter adjustments in real-time. This approach combining visual and haptic sensations can bring richer experiences and generate a realistic sensation of touching a grassy surface. Our objective is to enhance users' intuitive comprehension of haptic parameters through this innovative system.
\end{abstract}


\renewcommand\keywordsname{KEYWORDS}
\keywords{Midair Haptics, Haptic Display, Human-in-the-loop}
\begin{teaserfigure}
  \includegraphics[width=\textwidth]{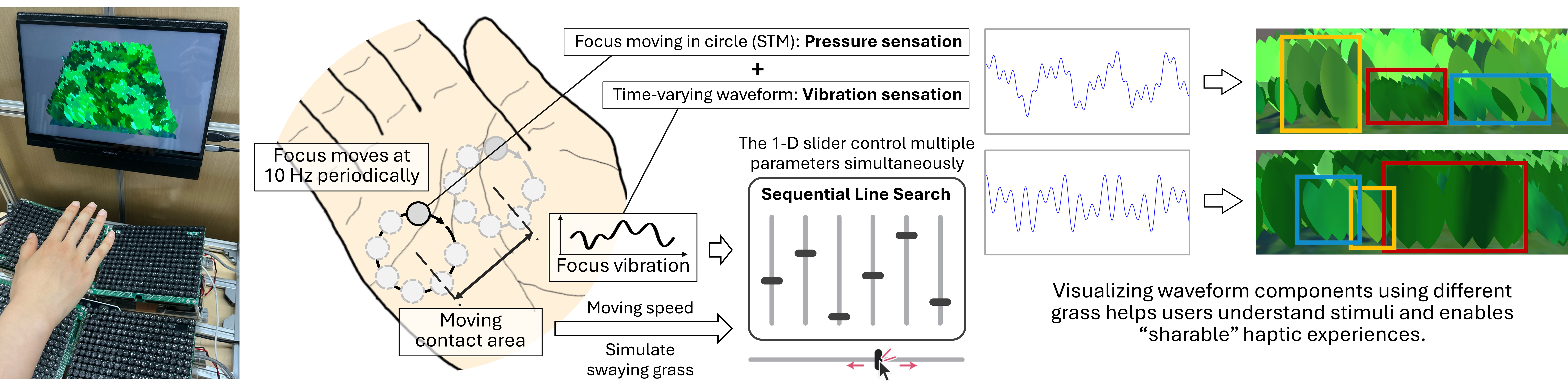}
  \caption{(Left) Our haptic presenting system. AUTDs produce haptic stimuli and the screen visualize haptic parameters using "Growing Grass". (Center) The presentation of haptic stimuli and the control of haptic parameters. (Right) Waveform change reflection on the visualized grass.}
  \label{fig:teaser}
\end{teaserfigure}


\maketitle

\section{INTRODUCTION}
Tactile interaction with virtual objects and surfaces is a crucial aspect of immersive technologies. Ultrasound-based midair haptic devices, such as the Airborne Ultrasound Tactile Display 3 (AUTD3) \cite{9392322}, have expanded the ways in which humans interact with virtual environments and objects, offering novel interaction experiences. These devices can present various pressure distributions to human skin, allowing users to feel different stimuli without the need to wear any contact device. Building on the AUTD3 platform, research efforts have focused on rendering haptic textures and integrating them with visual displays to create more realistic haptic display systems \cite{10.1145/3476122.3484849}.


However, designing suitable vibrations for various surfaces in the real world manually is a complex task that involves many parameters to adjust \cite{Incorporating-the-Perception}. This complexity makes it difficult for humans to perform the complex adjustment of multiple parameters during the optimization process simultaneously. Addressing this challenge requires a system capable of dynamically adapting to various vibration patterns and controlling multiple parameters concurrently. One possible solution is to employ methods successful in other fields, for example, Sequential Line Search (SLS) \cite{Sequential-Line-Search}, a human-in-the-loop method which allows users to control parameters through a one-dimensional (1-D) slider. Users can select the slider position and continue the iteration to finally achieve a desired feedback. By applying this method to manipulate mid-air ultrasound haptic parameters, we can simplify the process of designing suitable vibrations for various surfaces.

To enhance mid-air tactile sensations using Spatio-Temporal Modulation (STM), the focal point movement speed is optimized based on comprehensive user experiment \cite{frier2018using}. The distinguishability of mid-air ultrasound feedback as the desired feedback is investigated with varying temporal parameters for optimization, including amplitude-modulated (AM) frequencies, sinusoidal parameters, rhythms, and target metaphors \cite{10.1145/3613904.3642522}. These studies underscore the critical importance of optimizing stimulus parameters to desired sensations, whether by adjusting a single parameter or iteratively managing multiple parameters. However, the process of fine-tuning these parameters is both labor-intensive and requires a high degree of expertise, as users must conduct numerous experiments to comprehend the influence of each parameter on the resulting haptic sensation.

To address the need for a more intuitive and efficient system for understanding and controlling haptic parameters, integrating visual perception of roughness from images is proposed to aid in the design of mid-air haptic textures \cite{10.1145/3385955.3407927}. This approach leverages a visuo-haptic machine learning algorithm that correlates visual roughness ratings with tactile roughness ratings for various mid-air haptic stimuli. Building on this concept, RecHap tool which employs a neural network-based recommend system can suggest haptic patterns by exploring an encoded latent space \cite{RecHap2023}. RecHap also offers real-time visualization and manipulation of haptic sensations through an intuitive graphical user interface. While these methodologies effectively utilize large datasets to derive pre-trained weights, they need to establish target datasets as prior work and often lack flexibility in accommodating personalized customization needs.


Additionally, the combination of visual and haptic stimuli can enhance the sense of presence. By visualizing the tactile stimulus using the "Growing Grass" metaphor, we can create a more comprehensible and interactive experience, allowing users to understand the parameters more directly. Furthermore, we aim to create describable haptic features using the visualization so that the experience can be "shareable", helping users intuitively grasp haptic parameters.


\section{VISUAL-HAPTIC SYNCHRONOUS CONTROL SYSTEM}
\begin{figure}
	\centering
	\subfigure[]{
		\begin{minipage}[b]{\linewidth}
			\includegraphics[width=\linewidth]{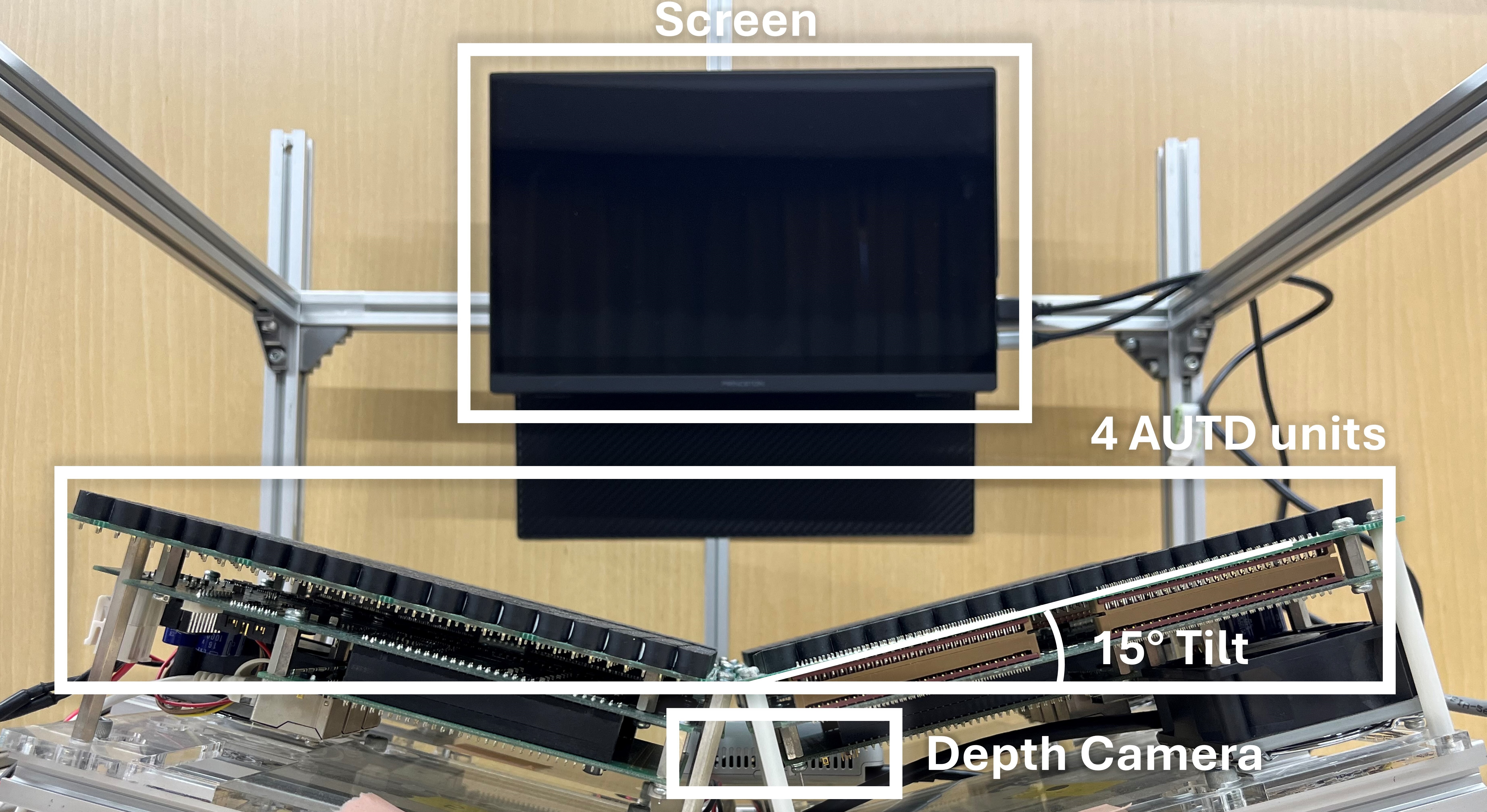} 
		\end{minipage}
		\label{fig:device}
	}
	\\ 
	\subfigure[]{
		\begin{minipage}[b]{0.59\linewidth}
			\includegraphics[width=1\linewidth]{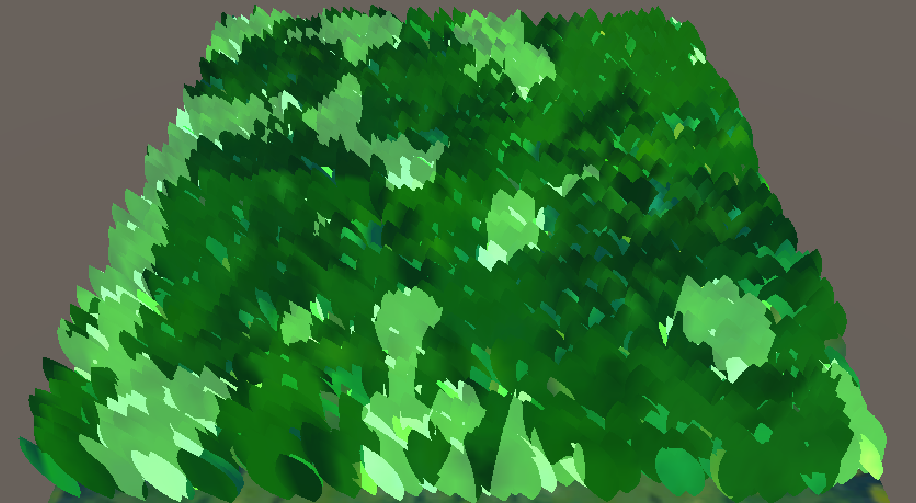} 
		\end{minipage}
		\label{fig:grass}
	}
    	\subfigure[]{
    		\begin{minipage}[b]{0.36\linewidth}
		 	\includegraphics[width=1\linewidth]{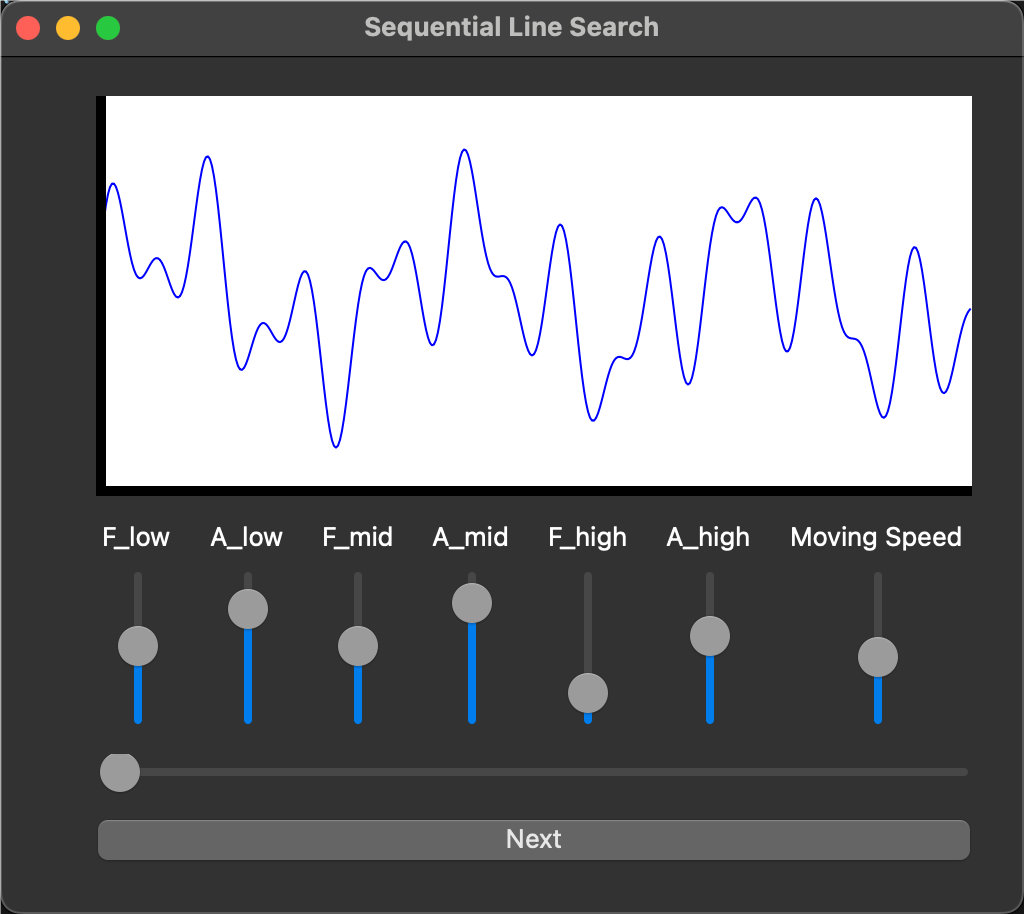}
    		\end{minipage}
		\label{fig:interface}
    	}
	
	\label{fig:configuration}
 \caption{(a) The system configuration. (b) The generated grass. (c) The user interface.}
\end{figure}

Our developed system shown in Fig. \ref{fig:device} consists of 4 AUTDs to present tactile stimuli, a screen to show the interface and visualization, and a depth camera (Intel Real Sense D435i). Four AUTD units are positioned at a 15-degree tilt. An AUTD unit has 249 ultrasound transducers driving at 40 kHz \cite{9392322}. The AUTDs present moving ultrasound focus to create a sense of contact area when the hand reaches a certain height.

\subsection{Presentation of the grass texture sensation}
The presentation method of the haptic sensation is shown in the left side of Fig. \ref{fig:teaser}. The system produces vibration and pressure sensations within a specific area using a periodically moving focus, employing a modulation technique known as Spatial Temporal Modulation (STM) \cite{frier2018using, 10.1145/3476122.3484849}. Ultrasound haptic stimuli can be weak due to sound wave limitations, STM method is proposed to enhance the static pressure sensation using a moving focus. What is more, STM can also be combined with vibrations, to create comprehensive touch experiences. The interaction area is fixed at a certain $x$-$y$ position, with a depth camera utilized solely to measure the distance from the system's origin to the user's hand. The radius of the focus moving circle in STM, representing the radius of the contact area, is set to 8 mm. This size is carefully chosen to avoid a point sensation from being too small and a too-prominent circular sensation from being too large. The STM frequency (repetition frequency of moving focus) is set at 10 Hz with a focus movement step size of 5 mm. These parameters are decided according to the system's shortest response time, to minimize system latency caused by real-time modulation waveform adjustments.

The waveform to produce vibration stimuli is composed of 3 sinusoidal waves representing low, medium, and high frequency components, respectively. The frequency ranges for these components are 10 - 30 Hz, 30 - 100 Hz and 100 - 300 Hz, respectively. Both the frequency and amplitude of each component can be dynamically adjusted using SLS or independently by users.

To simulate the swaying of grass, the system not only employs the moving focus to create a contact area but also allows this area to move laterally across the palm. The contact area oscillates periodically along a 30 mm straight path. Similar to the waveform adjustments, the moving frequency of the contact area (0.2 Hz - 1 Hz) can also be modified using SLS.


\subsection{Parameter controlling and visualization}
In this study, we control seven haptic parameters: the frequencies and amplitudes of the three waveform components, and the speed of the contact area movement. In this multi-parameter condition, the control can be much more complex, requiring many trials and a large cognitive load. SLS applies Bayesian Optimization method based on line-search oracle (that is the 1-D slider controlled by users) instead of traditional function value orcale, to simplify the process \cite{Sequential-Line-Search}. Using SLS, users can simultaneously and easily adjust these parameters to create the stimuli they desire or prefer. The user interface is shown in Fig. \ref{fig:interface}. Users can move the horizontal 1-D slider to change stimuli using SLS or directly move vertical sliders to change parameters independently to experience the impact brought by individual changes. The waveform is dynamically displayed on the interface in real-time.

To generate visual stimuli corresponding to the haptic feedback, we render three groups of grass in Unity3D, each representing one of the three waveform components (Fig. \ref{fig:grass}). The display region is fixed, with amplitudes indicated by the grass sizes and frequencies represented by the number of grass blades. The three groups of grass are colored differently to facilitate differentiation. We also implement a wind effect, where the rendered grass sways at different speeds according to the wind velocity, controlled by the moving speed of the haptic contact area. All parameter changes are reflected in the grass visualization in real-time, creating a coherent and immersive experience.

\section{CONCLUSION}
In this study, we proposed a method to simultaneously present ultrasound haptic stimuli and corresponding visual cues to enhance sensation and aid users in understanding haptic parameters. Leveraging the characteristics of vibration sensations, we visualized haptic stimuli using the "Growing Grass" metaphor. By applying a human-in-the-loop approach, we enabled users to easily control multiple parameters. We believe that our system not only provides a realistic experience but also facilitates the user understanding of tactile sensations.

\renewcommand{\acksname}{\textsc{ACKNOWLEDGMENTS}}
\begin{acks}
This work was supported in part by the JST SPRING, Grant Number JPMJSP2108.
\end{acks}

\renewcommand{\refname}{\textsc{REFERENCES}}
\bibliographystyle{ACM-Reference-Format}
\bibliography{ref}


\begin{thebibliography}{8}


\ifx \showCODEN    \undefined \def \showCODEN     #1{\unskip}     \fi
\ifx \showDOI      \undefined \def \showDOI       #1{#1}\fi
\ifx \showISBNx    \undefined \def \showISBNx     #1{\unskip}     \fi
\ifx \showISBNxiii \undefined \def \showISBNxiii  #1{\unskip}     \fi
\ifx \showISSN     \undefined \def \showISSN      #1{\unskip}     \fi
\ifx \showLCCN     \undefined \def \showLCCN      #1{\unskip}     \fi
\ifx \shownote     \undefined \def \shownote      #1{#1}          \fi
\ifx \showarticletitle \undefined \def \showarticletitle #1{#1}   \fi
\ifx \showURL      \undefined \def \showURL       {\relax}        \fi
\providecommand\bibfield[2]{#2}
\providecommand\bibinfo[2]{#2}
\providecommand\natexlab[1]{#1}
\providecommand\showeprint[2][]{arXiv:#2}

\bibitem[Suzuki et~al\mbox{.}(2021)]%
        {9392322}
\bibfield{author}{\bibinfo{person}{Shun Suzuki}, \bibinfo{person}{Seki Inoue}, \bibinfo{person}{Masahiro Fujiwara}, \bibinfo{person}{Yasutoshi Makino}, {and} \bibinfo{person}{Hiroyuki Shinoda}.} \bibinfo{year}{2021}\natexlab{}.
\newblock \showarticletitle{AUTD3: Scalable Airborne Ultrasound Tactile Display}.
\newblock \bibinfo{journal}{\emph{IEEE Transactions on Haptics}} \bibinfo{volume}{14}, \bibinfo{number}{4} (\bibinfo{year}{2021}), \bibinfo{pages}{740--749}.
\newblock
\urldef\tempurl%
\url{https://doi.org/10.1109/TOH.2021.3069976}
\showDOI{\tempurl}


\bibitem[Morisaki et~al\mbox{.}(2021)]%
        {10.1145/3476122.3484849}
\bibfield{author}{\bibinfo{person}{Tao Morisaki}, \bibinfo{person}{Masahiro Fujiwara}, \bibinfo{person}{Yasutoshi Makino}, {and} \bibinfo{person}{Hiroyuki Shinoda}.} \bibinfo{year}{2021}\natexlab{}.
\newblock \showarticletitle{Midair Haptic-Optic Display with Multi-Tactile Texture based on Presenting Vibration and Pressure Sensation by Ultrasound}. In \bibinfo{booktitle}{\emph{SIGGRAPH Asia 2021 Emerging Technologies}} (Tokyo, Japan) \emph{(\bibinfo{series}{SA '21})}. \bibinfo{publisher}{Association for Computing Machinery}, \bibinfo{address}{New York, NY, USA}, Article \bibinfo{articleno}{10}, \bibinfo{numpages}{2}~pages.
\newblock
\showISBNx{9781450386852}
\urldef\tempurl%
\url{https://doi.org/10.1145/3476122.3484849}
\showDOI{\tempurl}


\bibitem[Beattie et~al\mbox{.}(2020)]%
        {Incorporating-the-Perception}
\bibfield{author}{\bibinfo{person}{David Beattie}, \bibinfo{person}{William Frier}, \bibinfo{person}{Orestis Georgiou}, \bibinfo{person}{Benjamin Long}, {and} \bibinfo{person}{Damien Ablart}.} \bibinfo{year}{2020}\natexlab{}.
\newblock \showarticletitle{Incorporating the Perception of Visual Roughness into the Design of Mid-Air Haptic Textures}. In \bibinfo{booktitle}{\emph{ACM Symposium on Applied Perception 2020}} (Virtual Event, USA) \emph{(\bibinfo{series}{SAP '20})}. \bibinfo{publisher}{Association for Computing Machinery}, \bibinfo{address}{New York, NY, USA}, Article \bibinfo{articleno}{4}, \bibinfo{numpages}{10}~pages.
\newblock
\showISBNx{9781450376181}
\urldef\tempurl%
\url{https://doi.org/10.1145/3385955.3407927}
\showDOI{\tempurl}


\bibitem[Koyama et~al\mbox{.}(2017)]%
        {Sequential-Line-Search}
\bibfield{author}{\bibinfo{person}{Yuki Koyama}, \bibinfo{person}{Issei Sato}, \bibinfo{person}{Daisuke Sakamoto}, {and} \bibinfo{person}{Takeo Igarashi}.} \bibinfo{year}{2017}\natexlab{}.
\newblock \showarticletitle{Sequential Line Search for Efficient Visual Design Optimization by Crowds}.
\newblock \bibinfo{journal}{\emph{ACM Trans. Graph.}} \bibinfo{volume}{36}, \bibinfo{number}{4}, Article \bibinfo{articleno}{48} (\bibinfo{date}{jul} \bibinfo{year}{2017}), \bibinfo{numpages}{11}~pages.
\newblock
\showISSN{0730-0301}
\urldef\tempurl%
\url{https://doi.org/10.1145/3072959.3073598}
\showDOI{\tempurl}


\bibitem[Frier et~al\mbox{.}(2018)]%
        {frier2018using}
\bibfield{author}{\bibinfo{person}{William Frier}, \bibinfo{person}{Damien Ablart}, \bibinfo{person}{Jamie Chilles}, \bibinfo{person}{Benjamin Long}, \bibinfo{person}{Marcello Giordano}, \bibinfo{person}{Marianna Obrist}, {and} \bibinfo{person}{Sriram Subramanian}.} \bibinfo{year}{2018}\natexlab{}.
\newblock \showarticletitle{Using spatiotemporal modulation to draw tactile patterns in mid-air}. In \bibinfo{booktitle}{\emph{Haptics: Science, Technology, and Applications: 11th International Conference, EuroHaptics 2018, Pisa, Italy, June 13-16, 2018, Proceedings, Part I 11}}. Springer, \bibinfo{pages}{270--281}.
\newblock


\bibitem[Lim et~al\mbox{.}(2024)]%
        {10.1145/3613904.3642522}
\bibfield{author}{\bibinfo{person}{Chungman Lim}, \bibinfo{person}{Gunhyuk Park}, {and} \bibinfo{person}{Hasti Seifi}.} \bibinfo{year}{2024}\natexlab{}.
\newblock \showarticletitle{Designing Distinguishable Mid-Air Ultrasound Tactons with Temporal Parameters}. In \bibinfo{booktitle}{\emph{Proceedings of the CHI Conference on Human Factors in Computing Systems}} (<conf-loc>, <city>Honolulu</city>, <state>HI</state>, <country>USA</country>, </conf-loc>) \emph{(\bibinfo{series}{CHI '24})}. \bibinfo{publisher}{Association for Computing Machinery}, \bibinfo{address}{New York, NY, USA}, Article \bibinfo{articleno}{710}, \bibinfo{numpages}{18}~pages.
\newblock
\showISBNx{9798400703300}
\urldef\tempurl%
\url{https://doi.org/10.1145/3613904.3642522}
\showDOI{\tempurl}


\bibitem[Beattie et~al\mbox{.}(2020)]%
        {10.1145/3385955.3407927}
\bibfield{author}{\bibinfo{person}{David Beattie}, \bibinfo{person}{William Frier}, \bibinfo{person}{Orestis Georgiou}, \bibinfo{person}{Benjamin Long}, {and} \bibinfo{person}{Damien Ablart}.} \bibinfo{year}{2020}\natexlab{}.
\newblock \showarticletitle{Incorporating the Perception of Visual Roughness into the Design of Mid-Air Haptic Textures}. In \bibinfo{booktitle}{\emph{ACM Symposium on Applied Perception 2020}} (Virtual Event, USA) \emph{(\bibinfo{series}{SAP '20})}. \bibinfo{publisher}{Association for Computing Machinery}, \bibinfo{address}{New York, NY, USA}, Article \bibinfo{articleno}{4}, \bibinfo{numpages}{10}~pages.
\newblock
\showISBNx{9781450376181}
\urldef\tempurl%
\url{https://doi.org/10.1145/3385955.3407927}
\showDOI{\tempurl}


\bibitem[Theivendran et~al\mbox{.}(2023)]%
        {RecHap2023}
\bibfield{author}{\bibinfo{person}{Karthikan Theivendran}, \bibinfo{person}{Andy Wu}, \bibinfo{person}{William Frier}, {and} \bibinfo{person}{Oliver Schneider}.} \bibinfo{year}{2023}\natexlab{}.
\newblock \showarticletitle{RecHap: An Interactive Recommender System For Navigating a Large Number of Mid-Air Haptic Designs}.
\newblock \bibinfo{journal}{\emph{IEEE Transactions on Haptics}}  \bibinfo{volume}{PP}, Article \bibinfo{articleno}{Advance online publication} (\bibinfo{year}{2023}).
\newblock
\urldef\tempurl%
\url{https://doi.org/10.1109/TOH.2023.3276812}
\showDOI{\tempurl}


\end{thebibliography}

\end{document}